\pdfoutput=1
\RequirePackage[l2tabu, orthodox]{nag}

\documentclass[reqno,tbtags]{amsart}

\usepackage[notref,notcite]{showkeys} 

\usepackage{lmodern}
\usepackage[T1]{fontenc}
\usepackage[utf8]{inputenc}
\usepackage[english]{babel}
\usepackage{microtype} 

\usepackage{amsmath,amssymb,amsthm,mathrsfs,latexsym,mathtools,mathdots,csquotes,
booktabs,array,enumerate,tikz,bm,comment,hyperref}
\usepackage[centertableaux]{ytableau}

\usetikzlibrary{cd} 

\usetikzlibrary{arrows,positioning}
\usepackage[capitalize,noabbrev]{cleveref}

\usepackage{todonotes}
\presetkeys%
    {todonotes}%
    {inline,backgroundcolor=yellow}{}

\usepackage{graphicx}

\newtheorem{theorem}{Theorem}

\theoremstyle{definition}

\newtheorem{question}[theorem]{Question}
\newtheorem{problem}[theorem]{Problem}

\newcommand{\defin}[1]{\emph{#1}}

\newcommand{\setN}{\mathbb{N}}
\newcommand{\setZ}{\mathbb{Z}}

\newcommand{\setR}{\mathbb{R}}

\newcommand{\evec}{\mathbf{e}}

\title{NP-complete variants of some classical graph problems}

\date{
(\TeX\ \today{} \klockan)}

\begingroup
  \divide\count255 by 60
  \multiply\count255 by -60
  \advance\count255 by \time
  \ifnum \count255 < 10 \xdef\klockan{\the\count1.0\the\count255}
  \else\xdef\klockan{\the\count1.\the\count255}\fi
\endgroup

\author{Per Alexandersson}
\address{Dept. of Mathematics, Stockholm University, Sweden}
\email{per.w.alexandersson@gmail.com}


\begin{document}

\begin{abstract}
Some classical graph problems such as finding minimal spanning tree, shortest path or maximal flow
can be done efficiently. We describe slight variations of such problems which are shown to be NP-complete.
Our proofs use straightforward reduction from $3$-SAT.
\end{abstract}

\maketitle

\section{Introduction}

Graph theory is a rich source of interesting computational problems.
Consider the following classical decision problems regarding a graph $G$ with weighted edges:
\begin{itemize}
 \item determine if $G$ has a spanning tree with cost $\leq k$,
 \item determine if $G$ has a path from $u$ to $v$ in $G$ with cost $\leq k$,
 \item determine if $G$ has a flow from $u$ to $v$ in $G$ with value $\geq k$.
 \end{itemize}
These three questions can be answered efficiently, as there are algorithms for finding the minimal spanning tree (Prim's algorithm),
shortest path (Dijkstra), or maximal flow (Ford--Fulkerson). 
On the other hand, determining if $G$ has a Hamiltonian path, or if $G$ has a $k$-clique are
NP-complete decision problems, see \cite{Karp1972}.
Introducing additional restrictions on a problem with an efficient solution
may make the problem NP-complete. For example,
the problem of determining if a bipartite graph $(X,Y)$ has a perfect matching can be done in polynomial time.
D.~Plaisted consider the following variant, which he shows is NP-complete.
\begin{problem}[See \cite{Plaisted1980}]
Given a bipartite graph $(X,Y)$ and partitions $\mathcal{X}$, $\mathcal{Y}$ of $X$ and $Y$, respectively,
determine if there is a perfect matching such that there are no two edges $\{x,y\}$, $\{x',y'\}$
such that $x,x'$ are in the same block of $\mathcal{X}$ and $y,y'$ are in the same block of $\mathcal{Y}$.
\end{problem}

\bigskip 

In this note, we show that the following decision problems are NP-complete.
\begin{problem}[Vector-valued shortest path]\label{prob:vecPath}
 Let $G = (V,E)$ be a graph and let $w:E \to \setZ^N_{\geq 0}$ be a vector-valued weight on the edges.
 For a path $P = (v_1,v_2, \dotsc, v_\ell)$, let $w(P)$ be the vector $\sum_i w(v_i,v_{i+1})$.
 Let $u,v$ be vertices of $G$ and let $k \in \setR$.
 
 \begin{displayquote}
  Determine if there is a path $P$ in $G$ from $u$ to $v$ such that $|w(P)|\leq k$.
 \end{displayquote}
\end{problem}

\begin{problem}[Restricted spanning tree]\label{prob:spanningTree}
Let $G = (V,E)$ be a graph and $w:E \to \setZ_{\geq 0}$.
Let $F \subseteq E^2$ be a set of \emph{forbidden pairs} of edges, and let $k \in \setZ$.
\begin{displayquote}
  Determine if there is a spanning tree $T \subseteq E$ of $G$ such there is no pair of edges of $T$ in $F$,
  with cost $\leq k$.
\end{displayquote}
\end{problem}

\begin{problem}[All-or-nothing flow]\label{prob:Flow}
 
Let $N=(G,s,t,c)$ be a network with source $s$, sink $t$ and capacity $c$.
Furthermore, let $A$ be a subset of the edges of $G$.
A flow $f$ on $N$ is $A$-\emph{valid} if for all $(x,y) \in A$ we have either
$ f(x,y) = c(x,y) \text{ or } f(x,y) = 0$.
\begin{displayquote}
Determine if $N$ has an $A$-valid flow $f$ with $|f|\geq k$.
\end{displayquote}
\end{problem}

\subsection*{Acknowledgements}

This note was part of the requirements for a course in mathematical communication.
These proofs may serve as examples for students taking classes in graph theory or 
complexity theory.

\subsection{Preliminaries on complexity theory}

We assume that the reader is familiar with basic graph theory terminology. 
For a background on graph theory, we refer to \cite{Diestel2017}.

The expression ``NP'' refers to the class of decision problems solvable 
by a \emph{non-deterministic Turing machine} in \emph{polynomial time}.
It was shown by S.~Cook in 1971 \cite{Cook1971} that any such decision problem 
can be reduced to the so called \defin{$3$-SAT problem}.
This means that a problem which can be solved by a non-deterministic Turing machine,
can be phrased as an instances of $3$-SAT.
Furthermore, this translation can be carried out in polynomial time (polynomial in the size of the input).

\begin{problem}[3-SAT]
Let $E$ be a boolean expression, given as a conjunction of clauses, where each clause involves exactly three distinct literals.
A literal is either a boolean variable, or its negation. 
Determine if $E$ is satisfiable.
\end{problem}
For example, 
$
E = (x\vee y \vee \neg z) \wedge (\neg x\vee \neg y \vee \neg z) \wedge (x\vee \neg y \vee z)
$
is such a conjunction, and the assignment $x,z=\mathtt{true}$, $y=\mathtt{false}$ shows that $E$
can be satisfied. 

A problem class $C$ is said to be \defin{NP-complete} if any $3$-SAT problem can be 
reduced (in polynomial time) to an instance in $C$.
In 1972, R.~Karp presented a list of 21 NP-complete problems \cite{Karp1972} from various areas of mathematics,
thus highlighting the ubiquity of this notion.

To this date, Karp's list have been extended with a large variety of decision problems,
and we shall now extend this list even further.

\section{NP-complete variants of decision problems}

To prove that the problems listed in the introduction are NP-complete,
we shall describe reductions from $3$-SAT.
We use the boolean expression 
\begin{equation}\label{eq:booleanB}
B=
 (x \vee y \vee z) \wedge
 (\overline{x} \vee z \vee w) \wedge
 (\overline{x} \vee \overline{y} \vee w) \wedge
 (y \vee \overline{z} \vee \overline{w})
\end{equation}
for all our examples illustrating the reductions.

\subsection{Restricted spanning trees}

Given in instance of a $3$-SAT problem $B$ with $C$ clauses,
we construct a graph $G$ as in \eqref{eq:rst} as follows.

The graph $G=G(B)$ has one \emph{top vertex}, one \emph{bottom vertex}, and $(3+1)\cdot C$ auxiliary vertices.
The weights of all edges is $1$, except edges marked $\ast$, which has some fixed large weight $M$.
The three edges in each small ``block'' of $4$ vertices have been labeled according to a corresponding clause in $B$.
The forbidden set $F$ consists of all pairs of edges, $(a,\overline{a})$, 
where $a$ is a variable appearing in $B$.
\begin{equation}\label{eq:rst}
\begin{tikzpicture}[xscale=0.8,yscale=0.8]
\tikzset{vertNode/.style={circle,draw,minimum size=0.1cm,inner sep=0pt}}
\node[vertNode] (top) at ( 6.5, 1) {};
\node[vertNode] (bot) at ( 6.5, -2) {};
\begin{scope}[shift={(0,0)}]
    \node[vertNode] (v1a) at ( 1, 0) {};
    \node[vertNode] (v1b) at ( 2, 0) {};
    \node[vertNode] (v1c) at ( 3, 0) {};
    \node[vertNode] (v1out) at ( 2, -1) {};
\end{scope}
\begin{scope}[shift={(3,0)}]
    \node[vertNode] (v2a) at ( 1, 0) {};
    \node[vertNode] (v2b) at ( 2, 0) {};
    \node[vertNode] (v2c) at ( 3, 0) {};
    \node[vertNode] (v2out) at ( 2, -1) {};
\end{scope}
\begin{scope}[shift={(6,0)}]
    \node[vertNode] (v3a) at ( 1, 0) {};
    \node[vertNode] (v3b) at ( 2, 0) {};
    \node[vertNode] (v3c) at ( 3, 0) {};
    \node[vertNode] (v3out) at ( 2, -1) {};
\end{scope}
\begin{scope}[shift={(9,0)}]
    \node[vertNode] (v4a) at ( 1, 0) {};
    \node[vertNode] (v4b) at ( 2, 0) {};
    \node[vertNode] (v4c) at ( 3, 0) {};
    \node[vertNode] (v4out) at ( 2, -1) {};
\end{scope}

\draw[-] (v1out) to node[midway] {$x$} (v1a);
\draw[-] (v1out) to node[midway] {$y$} (v1b);
\draw[-] (v1out) to node[midway] {$z$} (v1c);

\draw[-] (v2out) to node[midway] {$\overline{x}$} (v2a);
\draw[-] (v2out) to node[midway] {$z$} (v2b);
\draw[-] (v2out) to node[midway] {$w$} (v2c);

\draw[-] (v3out) to node[midway] {$\overline{x}$} (v3a);
\draw[-] (v3out) to node[midway] {$\overline{y}$} (v3b);
\draw[-] (v3out) to node[midway] {$w$} (v3c);

\draw[-] (v4out) to node[midway] {$y$} (v4a);
\draw[-] (v4out) to node[midway] {$\overline{z}$} (v4b);
\draw[-] (v4out) to node[midway] {$\overline{w}$} (v4c);

\draw[-] (top) to node[midway] {} (v1a);
\draw[-] (top) to node[midway] {} (v1b);
\draw[-] (top) to node[midway] {} (v1c);
\draw[-] (top) to node[midway] {} (v2a);
\draw[-] (top) to node[midway] {} (v2b);
\draw[-] (top) to node[midway] {} (v2c);
\draw[-] (top) to node[midway] {} (v3a);
\draw[-] (top) to node[midway] {} (v3b);
\draw[-] (top) to node[midway] {} (v3c);
\draw[-] (top) to node[midway] {} (v4a);
\draw[-] (top) to node[midway] {} (v4b);
\draw[-] (top) to node[midway] {} (v4c);

\draw[thick,-] (bot) to node[midway] {$\ast$} (v1out);
\draw[thick,-] (bot) to node[midway] {$\ast$} (v2out);
\draw[thick,-] (bot) to node[midway] {$\ast$} (v3out);
\draw[thick,-] (bot) to node[midway] {$\ast$} (v4out);

\draw[-] (top) to node[midway] {} (bot);
\end{tikzpicture}
\end{equation}
In order for $G(B)$ to have a spanning tree which does not use any expensive (marked with $\ast$) edge,
we must be able to reach the $C$ vertices in the third layer, via some path from the top vertex.
This can only be done if and only if there is a truth-assignment of the boolean variables which makes $B$ true.
For the example in \eqref{eq:rst}, the assignment $x=\mathtt{false}$, $y=\mathtt{true}$, $z=\mathtt{true}$, $w=\mathtt{false}$
makes $B$ true, and there is a cheap minimal spanning tree, not using any $\ast$-edges:
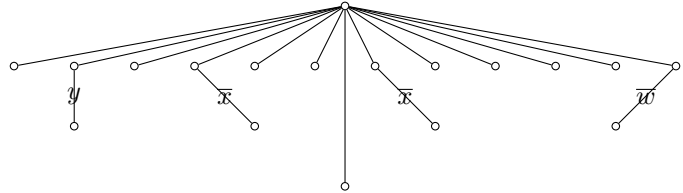
\[
\begin{tikzpicture}[xscale=0.8,yscale=0.8]
\tikzset{vertNode/.style={circle,draw,minimum size=0.1cm,inner sep=0pt}}
\node[vertNode] (top) at ( 6.5, 1) {};
\node[vertNode] (bot) at ( 6.5, -2) {};
\begin{scope}[shift={(0,0)}]
    \node[vertNode] (v1a) at ( 1, 0) {};
    \node[vertNode] (v1b) at ( 2, 0) {};
    \node[vertNode] (v1c) at ( 3, 0) {};
    \node[vertNode] (v1out) at ( 2, -1) {};
\end{scope}
\begin{scope}[shift={(3,0)}]
    \node[vertNode] (v2a) at ( 1, 0) {};
    \node[vertNode] (v2b) at ( 2, 0) {};
    \node[vertNode] (v2c) at ( 3, 0) {};
    \node[vertNode] (v2out) at ( 2, -1) {};
\end{scope}
\begin{scope}[shift={(6,0)}]
    \node[vertNode] (v3a) at ( 1, 0) {};
    \node[vertNode] (v3b) at ( 2, 0) {};
    \node[vertNode] (v3c) at ( 3, 0) {};
    \node[vertNode] (v3out) at ( 2, -1) {};
\end{scope}
\begin{scope}[shift={(9,0)}]
    \node[vertNode] (v4a) at ( 1, 0) {};
    \node[vertNode] (v4b) at ( 2, 0) {};
    \node[vertNode] (v4c) at ( 3, 0) {};
    \node[vertNode] (v4out) at ( 2, -1) {};
\end{scope}

\draw[-] (v1out) to node[midway] {$y$} (v1b);

\draw[-] (v2out) to node[midway] {$\overline{x}$} (v2a);

\draw[-] (v3out) to node[midway] {$\overline{x}$} (v3a);

\draw[-] (v4out) to node[midway] {$\overline{w}$} (v4c);

\draw[-] (top) to node[midway] {} (v1a);
\draw[-] (top) to node[midway] {} (v1b);
\draw[-] (top) to node[midway] {} (v1c);
\draw[-] (top) to node[midway] {} (v2a);
\draw[-] (top) to node[midway] {} (v2b);
\draw[-] (top) to node[midway] {} (v2c);
\draw[-] (top) to node[midway] {} (v3a);
\draw[-] (top) to node[midway] {} (v3b);
\draw[-] (top) to node[midway] {} (v3c);
\draw[-] (top) to node[midway] {} (v4a);
\draw[-] (top) to node[midway] {} (v4b);
\draw[-] (top) to node[midway] {} (v4c);
%

\draw[-] (top) to node[midway] {} (bot);
\end{tikzpicture}
\]
By picking $M>4C+1$, the question
\begin{displayquote}
Does $G(B)$ has a good spanning tree with cost $\leq M$ with no forbidden pairs in $F$? 
\end{displayquote}
is true if and only if $B$ has an assignment of variables that makes $B$ true.
We have thus showed that the restricted spanning tree problem is at least as hard as $3$-SAT.
Note that if we are given a spanning tree $T$ and a set $F$, and a number $k$,
we can easily check (in polynomial time) that $T$ is indeed a spanning tree, 
no pair of edges in $F$ appear in $T$, and the total cost does not exceed $k$.
It follows that our decision problem is NP-complete.

\subsection{All-or-nothing maximal flow}

For the next decision problem we use a construction which is similar to 
that of the restricted spanning tree problem.
Given a boolean expression $B$ with $C$ clauses and $V$ variables, 
we construct a network $(G,s,t,c)$ as follows,
see the figure in \cref{eq:flowExample}.
There is a top vertex $s$, the source, and a bottom vertex $t$, the sink.
The source is connected to $V$ vertices, $w_1,\dotsc,w_V$ positioned at level $2$.
Each $w_i$ is connected to two vertices on level $3$, labeled $x_i$ and $\overline{x}_i$.
These ``literal'' vertices are all connected to the vertex $l$, as well as 
the corresponding vertices in the $C$ clauses.

The dashed edges describe the elements in $A$.
We have the following capacities:
\begin{itemize}
 \item Any edge starting from the top $s$ has capacity $C$.
 \item The edge $(l,t)$ has capacity $VC-C$.
 \item All other edges ending at $t$ has capacity $1$.
 \item The edges in $A$ has capacity $C$.
 \item All the other edges are given a large capacity, say $VC$.
\end{itemize}

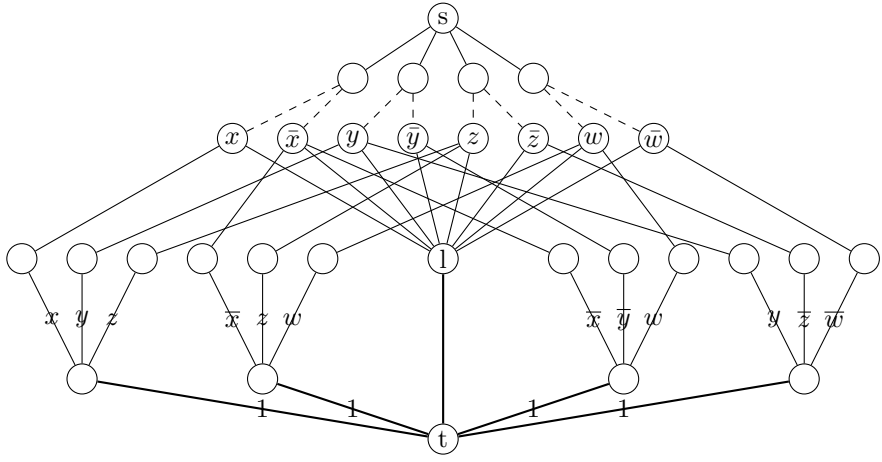
\begin{equation}\label{eq:flowExample}
\begin{tikzpicture}[xscale=0.8,yscale=0.8]
\tikzset{vertNode/.style={circle,draw,minimum size=0.4cm,inner sep=0pt}}
\tikzset{invisible/.style={minimum width=0mm,inner sep=0mm,outer sep=0mm}}
\node[vertNode] (top) at ( 8, 4) {s};
\node[vertNode] (bot) at ( 8, -3) {t};
\node[vertNode] (exess) at ( 8, 0) {l};


\begin{scope}[shift={(5.5,3)}]
\node[vertNode] (xc) at ( 1, 0) {};
\node[vertNode] (yc) at ( 2, 0) {};
\node[vertNode] (zc) at ( 3, 0) {};
\node[vertNode] (wc) at ( 4, 0) {};
\end{scope}

\begin{scope}[shift={(3.5,2)}]
\node[vertNode] (xx) at ( 1, 0) {$x$};
\node[vertNode] (xxb)at ( 2, 0) {$\bar{x}$};
\node[vertNode] (yy) at ( 3, 0) {$y$};
\node[vertNode] (yyb)at ( 4, 0) {$\bar{y}$};
\node[vertNode] (zz) at ( 5, 0) {$z$};
\node[vertNode] (zzb)at ( 6, 0) {$\bar{z}$};
\node[vertNode] (ww) at ( 7, 0) {$w$};
\node[vertNode] (wwb)at ( 8, 0) {$\bar{w}$};
\end{scope}

\begin{scope}[shift={(0,0)}]
    \node[vertNode] (v1a) at ( 1, 0) {};
    \node[vertNode] (v1b) at ( 2, 0) {};
    \node[vertNode] (v1c) at ( 3, 0) {};
    \node[vertNode] (v1out) at ( 2, -2) {};
\end{scope}
\begin{scope}[shift={(3,0)}]
    \node[vertNode] (v2a) at ( 1, 0) {};
    \node[vertNode] (v2b) at ( 2, 0) {};
    \node[vertNode] (v2c) at ( 3, 0) {};
    \node[vertNode] (v2out) at ( 2, -2) {};
\end{scope}
\begin{scope}[shift={(9,0)}]
    \node[vertNode] (v3a) at ( 1, 0) {};
    \node[vertNode] (v3b) at ( 2, 0) {};
    \node[vertNode] (v3c) at ( 3, 0) {};
    \node[vertNode] (v3out) at ( 2, -2) {};
\end{scope}
\begin{scope}[shift={(12,0)}]
    \node[vertNode] (v4a) at ( 1, 0) {};
    \node[vertNode] (v4b) at ( 2, 0) {};
    \node[vertNode] (v4c) at ( 3, 0) {};
    \node[vertNode] (v4out) at ( 2, -2) {};
\end{scope}

\draw[-] (v1out) to node[midway] {$x$} (v1a);
\draw[-] (v1out) to node[midway] {$y$} (v1b);
\draw[-] (v1out) to node[midway] {$z$} (v1c);

\draw[-] (v2out) to node[midway] {$\overline{x}$} (v2a);
\draw[-] (v2out) to node[midway] {$z$} (v2b);
\draw[-] (v2out) to node[midway] {$w$} (v2c);

\draw[-] (v3out) to node[midway] {$\overline{x}$} (v3a);
\draw[-] (v3out) to node[midway] {$\overline{y}$} (v3b);
\draw[-] (v3out) to node[midway] {$w$} (v3c);

\draw[-] (v4out) to node[midway] {$y$} (v4a);
\draw[-] (v4out) to node[midway] {$\overline{z}$} (v4b);
\draw[-] (v4out) to node[midway] {$\overline{w}$} (v4c);

\draw[-] (top) to node[midway] {} (xc);
\draw[-] (top) to node[midway] {} (yc);
\draw[-] (top) to node[midway] {} (zc);
\draw[-] (top) to node[midway] {} (wc);

\draw[dashed] (xc) to node[midway] {} (xx);
\draw[dashed] (xc) to node[midway] {} (xxb);
\draw[dashed] (yc) to node[midway] {} (yy);
\draw[dashed] (yc) to node[midway] {} (yyb);
\draw[dashed] (zc) to node[midway] {} (zz);
\draw[dashed] (zc) to node[midway] {} (zzb);
\draw[dashed] (wc) to node[midway] {} (ww);
\draw[dashed] (wc) to node[midway] {} (wwb);

\draw (exess) to node[midway] {} (xx);
\draw (exess) to node[midway] {} (xxb);
\draw (exess) to node[midway] {} (yy);
\draw (exess) to node[midway] {} (yyb);
\draw (exess) to node[midway] {} (zz);
\draw (exess) to node[midway] {} (zzb);
\draw (exess) to node[midway] {} (ww);
\draw (exess) to node[midway] {} (wwb);

\draw[-] (xx) to node[midway] {} (v1a);
\draw[-] (yy) to node[midway] {} (v1b);
\draw[-] (zz) to node[midway] {} (v1c);
\draw[-] (xxb) to node[midway] {} (v2a);
\draw[-] (zz) to node[midway] {} (v2b);
\draw[-] (ww) to node[midway] {} (v2c);
\draw[-] (xxb) to node[midway] {} (v3a);
\draw[-] (yyb) to node[midway] {} (v3b);
\draw[-] (ww) to node[midway] {} (v3c);
\draw[-] (yy) to node[midway] {} (v4a);
\draw[-] (zzb) to node[midway] {} (v4b);
\draw[-] (wwb) to node[midway] {} (v4c);

\draw[thick,-] (bot) to node[midway] {1} (v1out);
\draw[thick,-] (bot) to node[midway] {1} (v2out);
\draw[thick,-] (bot) to node[midway] {1} (v3out);
\draw[thick,-] (bot) to node[midway] {1} (v4out);

\draw[thick,-] (exess) to node[left] {$ $} (bot);


\end{tikzpicture}
\end{equation}
By construction, the restrictions imposed by $A$ imply that 
at most one of vertices $x$ and $\bar{x}$ has a non-zero flow (of size exactly $C$) leaving the vertex.

The edges leaving $s$ have total capacity $VC$, so this is an upper bound on 
the maximal flow in the network.
Furthermore, the flow $VC$ can only be obtained if
all edges flowing into $t$ are saturated,
as the total capacity of the edges adjacent to $t$ is also $VC$.
Hence, a maximal flow can exist only if every clause has some flow entering 
it (meaning at least one of the variables in the clause is true).
Hence, a maximal flow can only be obtained if the boolean expression is satisfiable.
On the other hand, if there is a way to satisfy $B$,
then we can find an edge in each clause (labeled with a literal being true) with flow $1$,
and let the remaining flow go via vertex $l$.
Hence, our network has a maximal flow with value $VC$ if and only if $B$ is satisfiable.

\subsection{Vector-valued path}

Given a boolean expression $B$ corresponding to an instance of $3$-SAT with $C$
clauses and $V$ variables, we construct a weighted graph $G$ on $4V+5C$ vertices as follows.
For each clause $c_i$, we have the vertices $\{a^i,w^i_1,w^i_2,w^i_3,b^i\}$,
and edges $(a^i,w^i_j)$, $(w^i_j,b^i)$ for $j=1,2,3$.
We also have the edges $(b^{i-1},a^{i})$ for all $i=2,\dotsc,C$.
This part of the graph is the \defin{clause subgraph}.

Furthermore, for each variable $x_i$, there are vertices $\{u_{in}^i,u^i,\overline{u}^i,u_{out}^i\}$,
and edges $(u_{in}^i,u^i)$, $(u^i,u_{out}^i)$, $(u_{in}^i,\overline{u}^i)$, $(\overline{u}^i,u_{out}^i)$
as well as the edges $(u_{out}^{i-1},u_{in}^{i})$ for all $i=2,\dotsc,V$.
This part of the graph is the \defin{variable subgraph}.
Finally, we have an edge $(u^V_{out},a^1)$ connecting the two subgraphs.

We shall now assign vector-valued weights with values in the $2V$-dimensional vector space 
$\setN^V \oplus \setN^V$. We let $\evec_i$ denote the $i$th unit vector in 
the first set of coordinates, and $\overline{\evec}_i$ the $i$th unit vector in the second 
set of coordinates.
The $V$ edges $(u^i_{in},u^i)$ and $(u^i_{in},\overline{u}^i)$ are given 
the weights $M \evec_i$ and $M \overline{\evec}_i$, respectively, 
where $M \gg 0$ is to be determined later.
The edge $(a^i,v^i_j)$ is given weight $\evec_\ell$ (or $\overline{\evec})$ if $x_\ell$
(or $\overline{x}_\ell$) is the variable at position $j$ in clause $i$, and 
the remaining edges in the graph are assigned weight $0$ in every coordinate.

We shall now argue that $B$ is satisfiable if and only if the question
\begin{displayquote}
Is there is a path $P$ from $u^1_{in}$ to $b^C$ 
with cost at most $\sqrt{VM^2 + C^3}$?
\end{displayquote}
has a positive answer.

\begin{figure}[!ht]
\[
\begin{tikzpicture}[xscale=1.2,yscale=1.2]
\tikzset{vertNode/.style={circle,draw,minimum size=0.1cm,inner sep=0pt}}

\begin{scope}[shift={(0,2)}]
    \node[vertNode] (w1a) at   ( 0, -0.5) {};
    \node[vertNode] (w1b) at   ( 0, 0.5) {};
    \node[vertNode] (w1in) at  (-0.5, 0) {};
    \node[vertNode] (w1out) at ( 0.5, 0) {};
\end{scope}
\begin{scope}[shift={(2,2)}]
    \node[vertNode] (w2a) at   ( 0, -0.5) {};
    \node[vertNode] (w2b) at   ( 0, 0.5) {};
    \node[vertNode] (w2in) at  (-0.5, 0) {};
    \node[vertNode] (w2out) at ( 0.5, 0) {};
\end{scope}
\begin{scope}[shift={(4,2)}]
    \node[vertNode] (w3a) at   ( 0, -0.5) {};
    \node[vertNode] (w3b) at   ( 0, 0.5) {};
    \node[vertNode] (w3in) at  (-0.5, 0) {};
    \node[vertNode] (w3out) at ( 0.5, 0) {};
\end{scope}
\begin{scope}[shift={(6,2)}]
    \node[vertNode] (w4a) at   ( 0, -0.5) {};
    \node[vertNode] (w4b) at   ( 0, 0.5) {};
    \node[vertNode] (w4in) at  (-0.5, 0) {};
    \node[vertNode] (w4out) at ( 0.5, 0) {$ $};
\end{scope}

\draw[-] (w1in) -- (w1a) -- (w1out);
\draw[-] (w1in) -- (w1b) -- (w1out);
\draw[-] (w2in) -- (w2a) -- (w2out);
\draw[-] (w2in) -- (w2b) -- (w2out);
\draw[-] (w3in) -- (w3a) -- (w3out);
\draw[-] (w3in) -- (w3b) -- (w3out);
\draw[-] (w4in) -- (w4a) -- (w4out);
\draw[-] (w4in) -- (w4b) -- (w4out);
\draw[-] (w1out) to (w2in);
\draw[-] (w2out) to (w3in);
\draw[-] (w3out) to (w4in);

\draw (w4out) to node[midway] {$x$} (w4a);
\draw (w4out) to node[midway] {$\overline x$} (w4b);
\draw (w3out) to node[midway] {$y$} (w3a);
\draw (w3out) to node[midway] {$\overline y$} (w3b);
\draw (w2out) to node[midway] {$z$} (w2a);
\draw (w2out) to node[midway] {$\overline z$} (w2b);
\draw (w1out) to node[midway] {$w$} (w1a);
\draw (w1out) to node[midway] {$\overline w$} (w1b);

\begin{scope}[shift={(0,0)}]
    \node[vertNode] (v1a) at   ( 0, -1) {};
    \node[vertNode] (v1b) at   ( 0, 0) {};
    \node[vertNode] (v1c) at   ( 0, 1) {};
    \node[vertNode] (v1in) at (-1, 0) {};
    \node[vertNode] (v1out) at ( 1, 0) {};
\end{scope}

\begin{scope}[shift={(2.5,0)}]
    \node[vertNode] (v2a) at   ( 0, -1) {};
    \node[vertNode] (v2b) at   ( 0, 0) {};
    \node[vertNode] (v2c) at   ( 0, 1) {};
    \node[vertNode] (v2in)  at (-1, 0) {};
    \node[vertNode] (v2out) at ( 1, 0) {};
\end{scope}

\begin{scope}[shift={(5,0)}]
    \node[vertNode] (v3a) at   ( 0, -1) {};
    \node[vertNode] (v3b) at   ( 0, 0) {};
    \node[vertNode] (v3c) at   ( 0, 1) {};
    \node[vertNode] (v3in)  at (-1, 0) {};
    \node[vertNode] (v3out) at ( 1, 0) {};
\end{scope}

\begin{scope}[shift={(7.5,0)}]
    \node[vertNode] (v4a) at   ( 0, -1) {};
    \node[vertNode] (v4b) at   ( 0, 0) {};
    \node[vertNode] (v4c) at   ( 0, 1) {};
    \node[vertNode] (v4in)  at (-1, 0) {};
    \node[vertNode] (v4out) at ( 1, 0) {$ $};
\end{scope}

\draw[-] (w1in) to (v1in);

\draw[-] (v1a) to (v1out);
\draw[-] (v1b) to (v1out);
\draw[-] (v1c) to (v1out);

\draw[-] (v2a) to (v2out);
\draw[-] (v2b) to (v2out);
\draw[-] (v2c) to (v2out);

\draw[-] (v3a) to (v3out);
\draw[-] (v3b) to (v3out);
\draw[-] (v3c) to (v3out);

\draw[-] (v4a) to (v4out);
\draw[-] (v4b) to (v4out);
\draw[-] (v4c) to (v4out);

\draw[-] (v1in) to node[midway] {$x$} (v1a);
\draw[-] (v1in) to node[midway] {$y$} (v1b);
\draw[-] (v1in) to node[midway] {$z$} (v1c);

\draw[-] (v2in) to node[midway] {$\overline{x}$} (v2a);
\draw[-] (v2in) to node[midway] {$z$} (v2b);
\draw[-] (v2in) to node[midway] {$w$} (v2c);

\draw[-] (v3in) to node[midway] {$\overline{x}$} (v3a);
\draw[-] (v3in) to node[midway] {$\overline{y}$} (v3b);
\draw[-] (v3in) to node[midway] {$w$} (v3c);

\draw[-] (v4in) to node[midway] {$y$} (v4a);
\draw[-] (v4in) to node[midway] {$\overline{z}$} (v4b);
\draw[-] (v4in) to node[midway] {$\overline{w}$} (v4c);

\draw[-] (v1out) to (v2in);
\draw[-] (v2out) to (v3in);
\draw[-] (v3out) to (v4in);

\end{tikzpicture}
\]
\caption{
Only the labeled edges have non-zero weight.
The path goes from the upper-rightmost vertex ($u^1_{in}$) to the lower right-most vertex ($b^C$).
}\label{fig:shortestPath}
\end{figure}
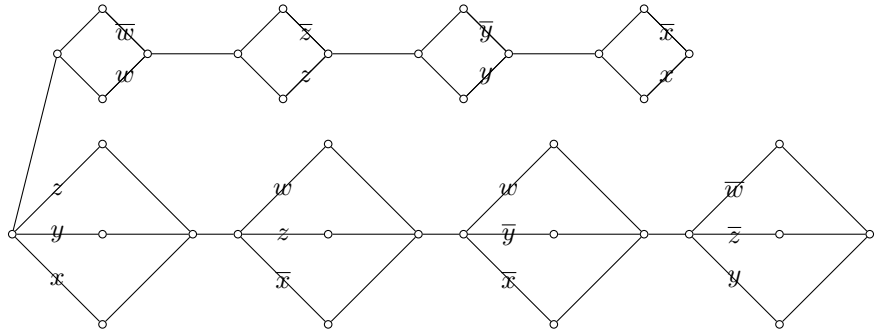

Any shortest path $P$ from $u^1_{in}$ to $b^C$ must use exactly 
one of the edges with weight $M\evec_i$ or $M\overline{\evec}_i$,
so at least half of the $2V$ available coordinates are non-zero in the cost of $P$.
Let us call these edges (and corresponding coordinates) in $P$ \defin{large edges}.
If there are no other edges in the path using the large coordinates,
the total cost squared is less than
\begin{equation}
 VM^2 + C^3, \tag{Satisfiable-upper-bound}
\end{equation}
as the maximal value in any of the remaining coordinates is $C$,
and there are at most $C$ such coordinates.

Suppose now that there is some additional edge using a large coordinate $\evec_i$ in $P$.
Then the total cost squared is at least 
\[
 (M+1)^2 + (V-1)M^2 > VM^2 + 2M, \tag{Unsatisfiable-lower-bound}
\]
the first term is the contribution from coordinate $\evec_i$, 
and there is a contribution from the $V-1$ remaining large coordinates $M$.
By choosing $M > \frac12 C^3$, we have that $VM^2 + 2M > VM^2 + C^3 $,
so there is a path with cost not exceeding 
$\sqrt{VM^2 + C^3}$
if and only if we can avoid having additional edges with non-zero value at a large coordinate.
Such a path can be found if and only if the corresponding $3$-SAT problem is satisfiable.

\begin{question}
 In the above setting, the length of the vector is allowed to depend on $V$, the number of variables.
What if the length is a fixed constant instead? 
Is the problem still in NP?
\end{question}

\bibliographystyle{alphaurl}
\bibliography{bibliography}

\end{document}